\documentclass[12pt]{article}
\usepackage[dvips]{graphicx}
\usepackage{psfrag}
\usepackage{epsf}
\usepackage[small]{caption}
\usepackage{subfigure}
\everymath{\displaystyle}
\newcommand{\bvec}[1]{\mbox{\boldmath $#1$}}
\begin{document}
\title{Macroscopic quantum tunneling of the Bose-Einstein 
condensate trapped in cylindrically symmetric potential}
\author{
\thanks{E-mail address : yasui@sci.osaka-cu.ac.jp} Yukinori YASUI,
\thanks{E-mail address : takaai@sci.osaka-cu.ac.jp} Takayuki TAKAAI and 
\thanks{E-mail address : ootsuka@sci.osaka-cu.ac.jp} Takayoshi OOTSUKA \\
{\small Department of Physics, Osaka City University, 
Sumiyoshiku, Osaka, Japan } }
\date{}
\maketitle
\begin{abstract}
We investigate the macroscopic quantum tunneling of the attractive 
Bose-Einstein condensate.
Within the effective Lagrangian framework,
we find bounce solutions and explicitly calculate the decay rate
of the condensate trapped in a cylindrically symmetric potential.
In particular, in the case where the number of condensed bosons
is slightly below a certain critical number,
we present a detailed analysis of the bounce solutions and 
discuss the approximations employed in our calculations.
The effects of finite temperatures and the shape of the 
trapping potential are evaluated.
\end{abstract}
\section{Introduction}

Macroscopic quantum tunneling is an interesting subject in
many areas of physical sciences including low-temperature 
physics, atomic physics and nuclear physics.
Recent realization of the Bose-Einstein condensate of trapped
alkali atoms may provide a good testing ground for the investigation
of this problem~\cite{D-G-P-S}.

In this paper we will discuss the macroscopic quantum tunneling
of the Bose-Einstein condensate with attractive interactions.
The dynamics of the condensate is successfully described by the
Gross-Pitaevskii (GP) equation~\cite{G, P}.
The s-wave scattering length $a$ entering the GP
equation can be positive
or negative, its sign and magnitude depending crucially on the 
details of the atom-atom interaction.
In the case of $^7$Li, the interaction is attractive and 
the scattering length is known to be 
$a=-1.45\pm 0.04~{\rm nm}$~\cite{M-S-H-V, A-Mc-S-H}. 
The attractive interaction
causes the condensate to collapse upon itself.
When the trapping potential is included, however,
the destabilizing influence of the interaction is
balanced by the zero-point kinetic energy,
thereby allowing a metastable condensate to form~\cite{P-M-C-L-Z, S, P-S-R}.
P\'erez-Garc\'{\i}a et al.~\cite{P-M-C-L-Z}
have investigated the GP equation by using
a time-dependent variational ansatz for the condensate wave function.
Their results reproduce quite accurately the low energy excitation
spectrum of the condensate
obtained by numerical simulations of the GP equation.
We will apply this variational technique to the macroscopic
tunneling of the metastable condensate of $^7$Li.
When the trapping is spherically symmetric,
Ueda and Leggette have evaluated the tunneling decay rate
at zero temperature~\cite{U-L} (see also~\cite{S, H-M-D-B-B}).
In this paper we develop their analysis and explicitly write down
the decay rate in the case of a cylindrically symmetric trapping
potential and further finite temperatures.

In Section 2, according to~\cite{P-M-C-L-Z}, we derive an effective
Lagrangian describing the Bose-Einstein
condensate of $^7$Li, and summarize the data of the ground
state energy that we shall need in the calculations of tunneling.
In Section 3, we present a detailed analysis of bounce solutions.
Using the effective Lagrangian and with the help of numerical
simulations, we find the bounce solutions.
We next consider the special situation, where the number of 
condensed bosons is slightly below a certain critical number.
Then the effective Lagrangian reduces to a simple one-dimensional
Lagrangian by appropriate approximations.
We present an analytic solution for the bounce within this situation,
and explicitly calculate the decay rate of the metastable condensate.
We also evaluate the decay rate at finite temperatures and
predict a critical temperature, where the rate crosses over
from quantum tunneling to thermal hopping.
Section 4 is devoted to the summary of our findings.

\section{Model}
We consider gases of $^7$Li atoms trapped
in a cylindrically symmetric harmonic potential
\begin{eqnarray}
 V(x,y,z) = \frac{1}{2} m \nu ^2 (x^2 + y^2 + \lambda ^2 z^2),
\end{eqnarray}
where $\lambda$ represents the asymmetry parameter of the trapping potential.
The dynamics of the condensate is described 
by the GP Lagrangian
\begin{eqnarray}
& & {\cal L} = 
   \frac{i \hbar}{2} \left( \psi \frac{\partial {\psi ^*}}{\partial t}
             - \frac{\partial \psi}{\partial t}\psi ^* \right)
             -\frac{\hbar ^2}{2m} |\bigtriangledown \psi |^2 -V|\psi |^2 
             -\frac{2\pi \hbar ^2 a}{m} |\psi |^4 \label{L}.
\end{eqnarray}

In order to obtain the evolution of the condensate wave function,
we assume the Gaussian form for the wave function according to~\cite{P-M-C-L-Z}:
\begin{eqnarray}
  \psi (x,y,z,t) = A(t) \prod _{a=x,y,z} 
  \exp \left[
                -\frac{(x_a - \eta _a (t))^2}{2 W_a (t) ^2} + i x_a \alpha _a (t)
                +i x_a ^2 \beta _a (t) \right] \label{Psi}.
\end{eqnarray}
This trial function includes the time-dependent variational parameters, 
$\bvec{\eta} = (\eta _x, \eta _y, \eta _z)$ (center coordinate), $\bvec{
W} = (W_x, W_y, W_z)$(width) and the phase parameters
$\bvec{\alpha} = (\alpha _x, \alpha _y , \alpha _z) $, $\bvec{\beta} =
(\beta _x, \beta _y, \beta _z)$ which correspond to the canonically conjugate
``momentums'' to $\bvec{\eta}$ and $\bvec{W}$.
The wave function $\psi$ is normalized by the number of condensed bosons
$N = \int |\psi |^2 d^3 x$, so that the parameter $A$ (amplitude) is
given by
\begin{eqnarray}
 A = \frac{1}{\pi ^{3/4}} \sqrt{\frac{N}{W_x W_y W_z}}.
\end{eqnarray}
Substituting (\ref{Psi}) into (\ref{L}) and 
further integrating the GP Lagrangian over space coordinates, 
one obtains an effective quantum mechanical
Lagrangian
\begin{eqnarray}\label{Leff}
 {\cal L}_{eff} = \sum _{a=x,y,z} (p_a \dot{\eta} _a + K_a \dot{W} _a)
  - {\cal H}_{eff} (\eta _a, W_a, p_a, K_a),
\end{eqnarray}
where $p_a$ and $K_a$ are the canonically conjugate momentums to $\eta
_a$ and $W_a$ defined by
\begin{eqnarray}
 p_a &=& \hbar N (\alpha _a + 2 \eta _a \beta _a), \\
 K_a &=& \hbar N \beta _a W_a.
\end{eqnarray}
The Hamiltonian ${\cal H}_{eff} = {\cal H}_0 + {\cal H}$ consists of two parts: the first
part ${\cal H}_0$ simply describes the harmonic oscillation of the center of the
condensate
\begin{eqnarray}
 {\cal H}_0 = \sum_{a=x,y,z} \frac{1}{2mN} p_a ^2
  + \frac{Nm\nu^2}{2} (\eta _x ^2 + \eta _y ^2 + \lambda ^2 \eta _z ^2),
\end{eqnarray}
and the remaining part ${\cal H}$ describes the evolution
of the widths of the condensate
\begin{eqnarray}
 {\cal H} = \sum _{a=x,y,z} \frac{1}{mN} K_a ^2 + \hat{U}(\bvec{W})
\end{eqnarray}
with
\begin{eqnarray}
  \hat{U}(\bvec{W}) &=& \frac{mN\nu ^2}{4}  (W_x^2 + W_y^2 + \lambda ^2W_z^2)
  +\frac{\hbar ^2 N}{4m} \left(\frac{1}{W_x^2} + \frac{1}{W_y^2} 
                           + \frac{1}{W_z^2}\right) 
   + \frac{a \hbar^2 N^2}{\sqrt{2 \pi}m}\frac{1}{W_xW_yW_z}.\label{uhat}
\end{eqnarray}
It is convenient to introduce the scales characterizing the trapping
potential : (a) length scale $a_0 = \sqrt{\hbar/{m\nu}}$, (b)
energy scale $e_0 = \hbar \nu /2$, (c) time scale $\nu ^{-1}$.
By using these units we define dimensionless quantities, $\bvec{\xi}
= a_0^{-1} \bvec{\eta}, \bvec{X} = a_0^{-1} \bvec{W}$ and $\tau  = \nu t$.
Then the Lagrangian (\ref{Leff}) is rescaled as follows :
\begin{equation}
 L_{eff} = e_0 ^{-1} {\cal L}_{eff} = L_0 + L,
\end{equation}
where
\begin{equation}
 L_0 = N \left(\frac{d\bvec{\xi}}{d\tau}\right)^2
  - N ( \xi _x^2 + \xi _y^2 + \lambda ^2 \xi _z^2),
\end{equation}
and
\begin{eqnarray}
 &L = \frac{N}{2} \left(\frac{d\bvec{X}}{d\tau}\right)^2
  - N U(\bvec{X}), \label{LN}\\
 &U(\bvec{X}) =  \frac{1}{2} (X^2 + Y^2 + \lambda ^2Z^2) +\frac{1}{2} 
  \left( \frac{1}{X^2} + \frac{1}{Y^2} + \frac{1}{Z^2}\right) 
  + \frac{P}{XYZ}\label{ux}
\end{eqnarray}
 with $P = \sqrt{2/\pi}  Na/a_0 < 0$.
We now focus our attention on the ground state energy of the condensed
Bose system. Under the present analysis, the ground state energy can be
calculated by finding the critical points of $U$ and the eigenvalues of
the Hessian matrix 
$H_{ab} = \partial ^2 U/\partial X_a \partial X_b$
evaluated on the critical points.
\newpage

\begin{figure}[h]
\begin{center}
  \psfrag{0.001}[]{\scriptsize $0.001$}
  \psfrag{0.01}[]{\scriptsize $0.01$}
  \psfrag{0.1}[]{\scriptsize $0.1$}
  \psfrag{1}[]{\scriptsize $1$}
  \psfrag{10}[]{\scriptsize $10$}
  \psfrag{100}[]{\scriptsize $100$}
  \psfrag{20}[]{\scriptsize $20$}
  \psfrag{40}[]{\scriptsize $40$}
  \psfrag{60}[]{\scriptsize $60$}
  \psfrag{80}[]{\scriptsize $80$}
  \psfrag{100}[]{\scriptsize $100$}
  \psfrag{1000}[]{\scriptsize $1000$}
  \psfrag{-0.8}[]{\scriptsize $-0.8$}
  \psfrag{-0.6}[]{\scriptsize $-0.6$}
  \psfrag{-0.4}[]{\scriptsize $-0.4$}
  \psfrag{-0.2}[]{\scriptsize $-0.2$}
  \psfrag{0}[]{\scriptsize $0$}
  \psfrag{1}[]{\scriptsize $1$}
  \psfrag{0.8}[]{\scriptsize $0.8$}
  \psfrag{0.6}[]{\scriptsize $0.6$}
  \psfrag{0.4}[]{\scriptsize $0.4$}
  \psfrag{0.2}[]{\scriptsize $0.2$}
 \psfrag{lambda}[]{$\lambda$}
 \psfrag{xc}[]{$X^* = Y^*$}
 \psfrag{zc}[]{$Z^*$}
 \psfrag{pc}[]{$P^*$}
 \includegraphics[width=10cm,height=10cm,keepaspectratio]{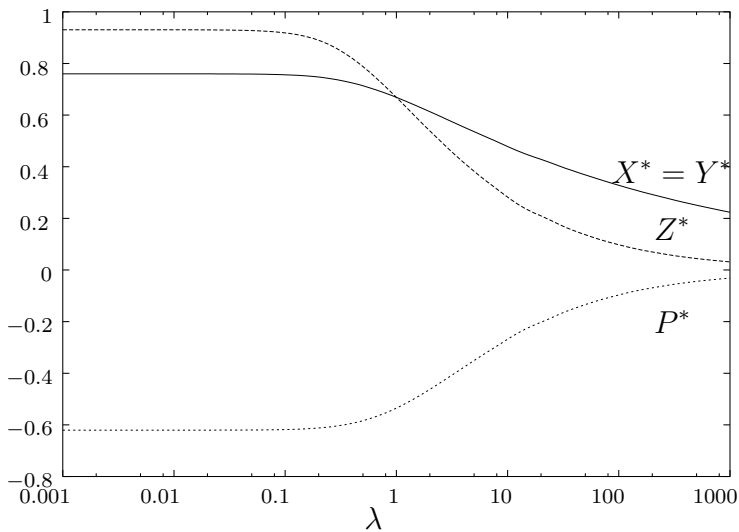}
 \caption{$\lambda$-dependence of $P^*$ and $\bvec{X}^* = (X^*, Y^*, Z^*)$.}
 \label{fig:l-xcyczcpc}
 \end{center}
\end{figure}

\vspace{12pt}
\begin{flushleft}
{\bf Critical points}
\end{flushleft}
The critical points are given by
the solutions to
\begin{eqnarray}
 &X = Y, \label{crit1}\\
 &\frac{1}{Z^4} + \frac{P}{X^2 Z^3} = \lambda ^2, \label{crit2} \\
 &\frac{1}{X^4} + \frac{P}{X^4 Z} = 1\label{crit3}.
\end{eqnarray}
The solutions are classified by the critical value $P^*$ of the
parameter $P$; when $|P| > |P^*|$, there are no critical points. When
$|P|<|P^*|$, there are two critical points, one stable (Morse index = 0)
and the other unstable (Morse index = 1)~\cite{P-M-C-L-Z}.
The critical value $P^*$ satisfies in addition to the equations
(\ref{crit1}) (\ref{crit2}) (\ref{crit3}) also 
\begin{eqnarray}
 \frac{P}{X^2 Z^3} + \frac{1}{2} \frac{P^2}{X^6 Z^4} = 4\lambda ^2,
\end{eqnarray}
which can be derived from the condition $\epsilon _T =0$ (see (\ref{lambda-t})).
Thus, $P^*$ and the corresponding coordinate $\bvec{X}^* =
(X^*, Y^*, Z^*)$ are uniquely determined as a function of the
asymmetry parameter $\lambda$.
Indeed we have $P^*=-4/5^{5/4}$, $\bvec{X}^*=5^{-1/4}(1,1,1)$
for $\lambda=1$,
and general solutions are provided in Fig.\ref{fig:l-xcyczcpc}.
It should be noticed that for $P \to P^*$ the stable
critical point $\bvec{X}_s$ and the unstable critical point 
$\bvec{X}_u$ take the following asymptotic forms:
\begin{eqnarray}\label{crit}
 \bvec{X}_{s, \,u}
=\bvec{X}^* \pm k(1-P/P^*)^{1/2} \bvec{E} + {\cal O}(1-P/P^*),
\end{eqnarray}
where $\bvec{E}=(-P^*_{32}, -P^*_{32}, 1)$ and the coefficient $k$
is given by
\begin{eqnarray}
 k = \sqrt{\frac23} \left( \frac{P^*_{21}(2P^*_{41}-1)}
                     {2 \lambda^2 (1-P^*_{41})-P^*_{23}}
                    \right)^\frac12
\end{eqnarray}
with $P^*_{ij}= P^*/4(X^*)^i(Z^*)^j$.


\vspace{12pt}
\begin{flushleft}
{\bf Eigenvalues of Hessian matrix}
\end{flushleft}
For the eigenvalue problem of the Hessian matrix evaluated on the
critical points
\begin{eqnarray}
 H \bvec{e}_A = \epsilon _A ^2 \bvec{e}_A\quad (A=T,N,B),
\end{eqnarray}
we have the following results~\cite{P-M-C-L-Z}:
\begin{enumerate}
 \item[(a)] $T$-direction
           \begin{eqnarray}
                \epsilon _T ^2 &=& 2\left(\lambda ^2 + 1 - P_{23} - 
\sqrt{8(P_{32})^2 + (1-\lambda ^2 + P_{23})^2} \right),\label{lambda-t}\\
                 \bvec{e}_T &=& \frac{1}{\triangle_T} (T, T, P_{32}),\label{eT}\\
                {\mathrm with}\quad T &=& \frac{1}{4}\left(-\lambda ^2 + 1 + P_{23} 
   - \sqrt{8(P_{32})^2 + (1-\lambda ^2 + P_{23})^2} \right). \label{T}
           \end{eqnarray}
 \item[(b)] $N$-direction
           \begin{eqnarray}
                \epsilon _N^2 &=& 2\left(\lambda ^2 + 1 - P_{23} +
 \sqrt{8(P_{32})^2 + (1-\lambda ^2 + P_{23})^2} \right),\label{lambda-n}\\
                 \bvec{e}_N &=& \frac{1}{\triangle_N} (N, N, P_{32}),\\
                {\mathrm with}\quad N &=& \frac{1}{4}\left(-\lambda ^2 + 1 
+ P_{23} + \sqrt{8(P_{32})^2 + (1-\lambda ^2 + P_{23})^2} \right).
           \end{eqnarray}
 \item[(c)] $B$-direction
                        \begin{eqnarray}
                         \epsilon _B^2 = 4(1-2P_{41}),\label{lambda-b}\\
                          \bvec{e}_B = \frac{1}{\sqrt{2}} (1,-1,0).
                        \end{eqnarray}
\end{enumerate}
Here we used the notation
\begin{equation}
 P_{ij} = \frac{P}{4 X^i Z^j}\quad (|P|\le|P^*|) \label{pij},
\end{equation}
and
\begin{equation}
\triangle_{T, \, N} ^2 = 2(P_{32})^2 + 
 \frac{1}{4} \left[ (P_{23} + 1 - \lambda ^2)^2 \mp (P_{23} + 1 
                          - \lambda ^2)\sqrt{8(P_{32})^2 + (1-\lambda ^2 + P_{23})^2}                                   \right]
\end{equation}
by the normalization $\bvec{e}_A \cdot \bvec{e}_B = \delta _{AB}$.
It should be noticed that the eigenvalue $\epsilon _T ^2$ is positive
(negative) for the stable (unstable) critical point and the other
eigenvalues are all positive. These results imply the following ground state
energy
\begin{equation}
 e_0^{-1}E_g = NU( \bvec{X}_s) 
+ (2+\lambda) + \sum_{A=T,N,B} \epsilon _A (\bvec{X}_s).\label{eg}
\end{equation}
The first term is the potential energy evaluated on the stable critical
point $\bvec{X}_s$, and the second-third terms represent
the zero-point energy coming from collective excitations of the
condensate. 

\section{Macroscopic quantum tunneling}
In this section we argue the macroscopic quantum tunneling of the
Bose condensate using the Lagrangian (\ref{LN}).
The stable critical point $\bvec{X}_s$ of the potential $U(\bvec{X})$
represents a metastable condensate since the parameter $P$ in $U(\bvec{X})$
is negative, and so the
ground state energy will have an (exponentially small) imaginary part in
addition to (\ref{eg}) if we take account of the tunneling. 
The decay rate of the metastable condensate is determined from
\begin{equation}
 \Gamma= \frac{2}{\hbar} {\rm Im} E_{\rm g}.
\end{equation}
We will calculate the decay rate by using the WKB approximation.
Since the Lagrangian (\ref{LN}) includes a macroscopic quantity $N$
representing the number of condensed bosons,
we must be careful for the choice of a small parameter $h$ controlling
the validity of the WKB approximation. 
The precise value of $h$ is given by (\ref{h}), and 
the decay rate is of the form
\begin{equation}\label{Gamma}
 \Gamma \simeq A  \exp \left(-\frac{S_{\rm cl}}{h}\right),
\end{equation}
where $S_{\rm cl}$ is the Euclidean action evaluated at bounce solution 
and $A$ the square root of the determinant of the second
variation around the bounce solution, with the zero - mode removed.

\begin{figure}[htbp]
 \begin{center}
  \subfigure[$\lambda =0.5,P=-0.5$]{%
  \psfrag{x}[]{\small $X=Y$}
  \psfrag{z}[]{\small $Z$}
  \psfrag{0.3}[]{\scriptsize $0.3$}
  \psfrag{0.4}[]{\scriptsize $0.4$}
  \psfrag{0.5}[]{\scriptsize $0.5$}
  \psfrag{0.6}[]{\scriptsize $0.6$}
  \psfrag{0.7}[]{\scriptsize $0.7$}
  \psfrag{0.8}[]{\scriptsize $0.8$}
  \psfrag{0.9}[]{\scriptsize $0.9$}
  \psfrag{1}[]{\scriptsize $1$}
  \psfrag{xs}[]{\small $\bvec{X}_s$}
  \psfrag{xb}[]{\small $\bvec{X}_t$}
  \includegraphics[width=7cm,height=7cm,keepaspectratio]{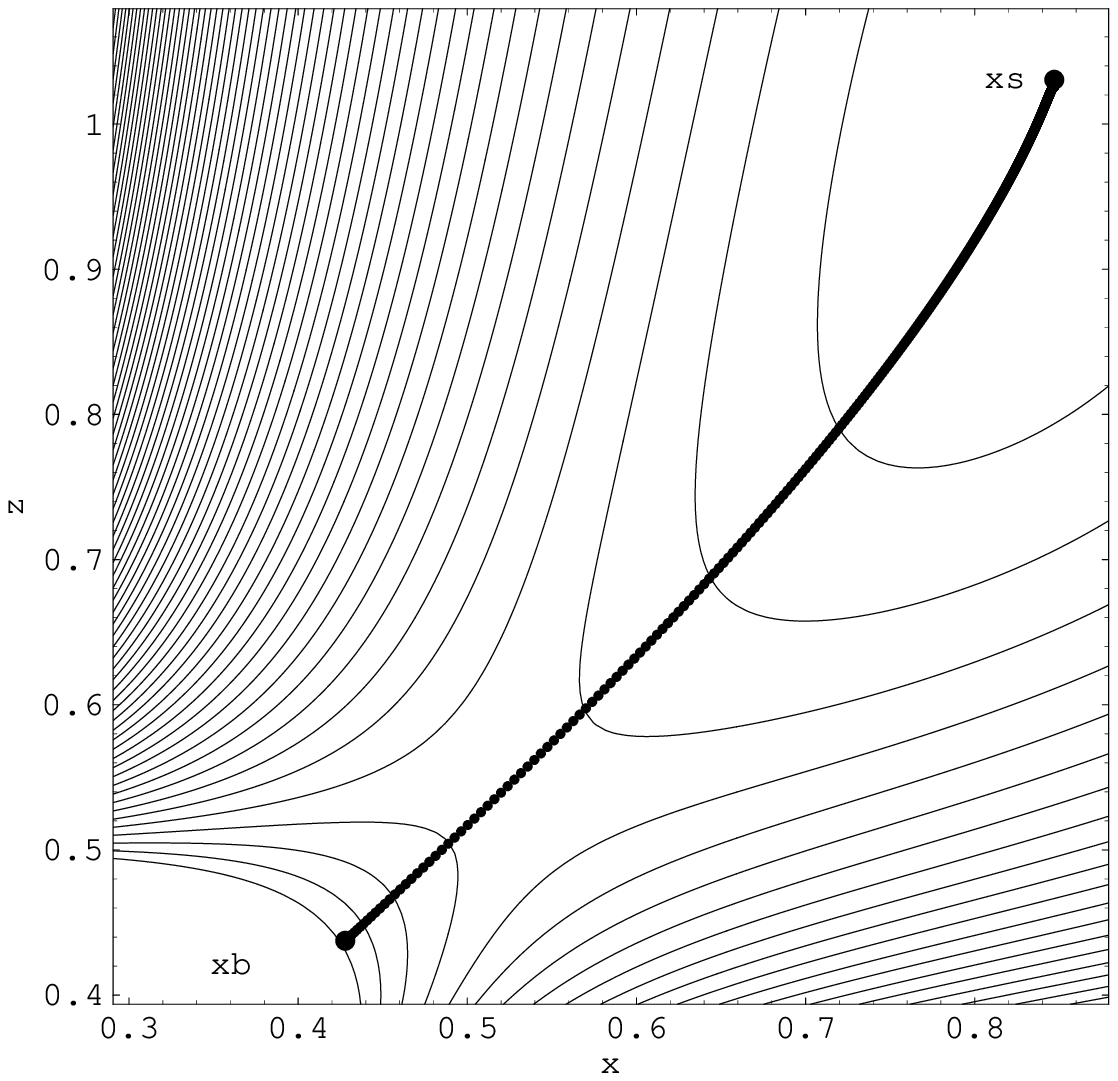}
  }
  \subfigure[$\lambda =2.0, P=-0.4$]{
  \psfrag{x}[]{\small $X=Y$}
  \psfrag{z}[]{\small $Z$}
  \psfrag{0.3}[]{\scriptsize $0.3$}
  \psfrag{0.4}[]{\scriptsize $0.4$}
  \psfrag{0.5}[]{\scriptsize $0.5$}
  \psfrag{0.6}[]{\scriptsize $0.6$}
  \psfrag{0.7}[]{\scriptsize $0.7$}
  \psfrag{0.8}[]{\scriptsize $0.8$}
  \psfrag{0.35}[]{}
  \psfrag{0.45}[]{}
  \psfrag{0.55}[]{}
  \psfrag{xs}[]{\small $\bvec{X}_s$}
  \psfrag{xb}[]{\small $\bvec{X}_t$}
  \includegraphics[width=7cm,height=7cm,keepaspectratio]{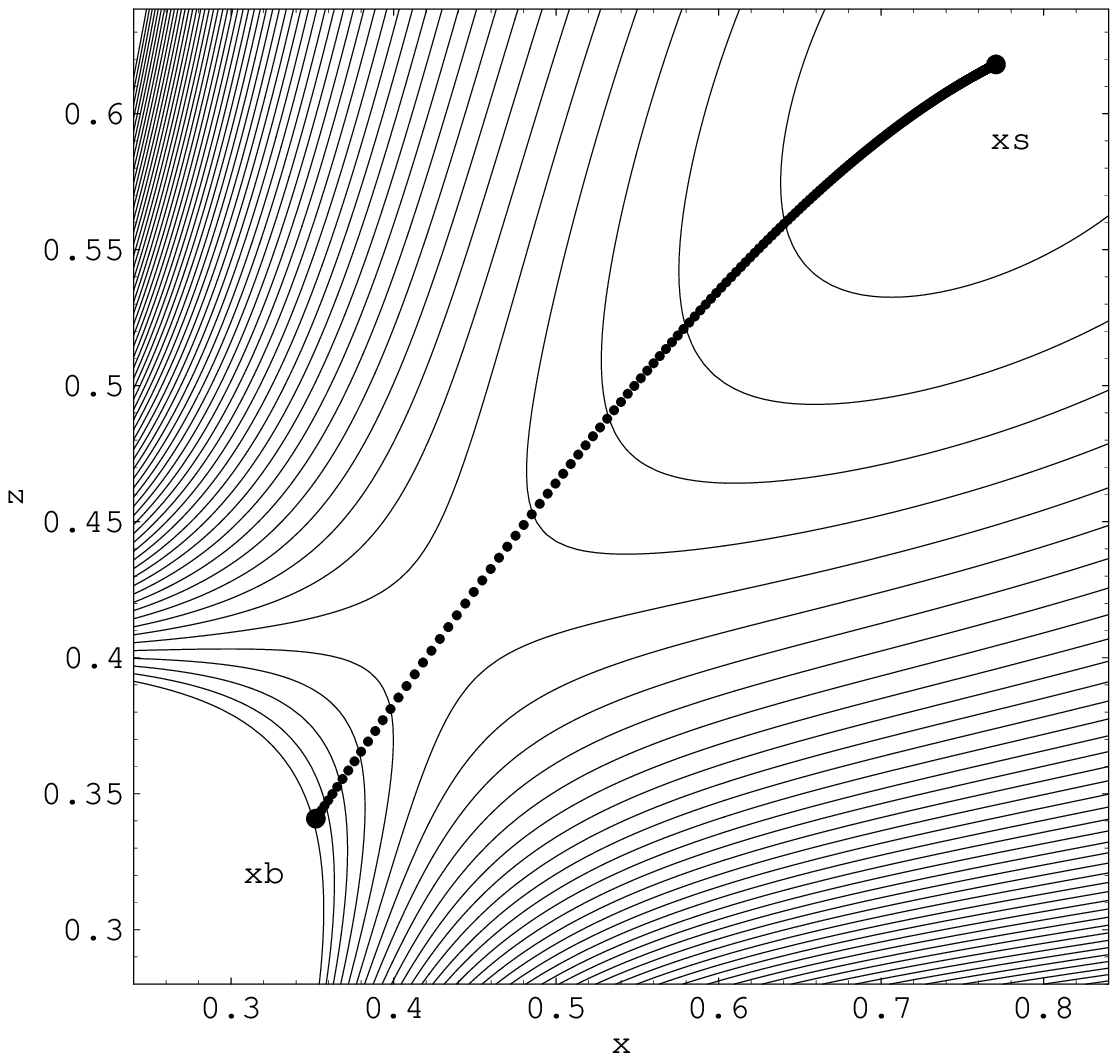}}
  \caption{Behavior of the bounce solution. The bold-faced curve
  connecting the two points, $\bvec{X}_s$ (stable critical point) and 
  $\bvec{X}_t$ (turning point) corresponds to the bounce solution. 
  The solid curves represent the contours of the potential
  $U(\bvec{X})$.
  Parameter values : $\delta = 1-P/P^* = 0.144$ for (a) and $\delta = 0.135$ 
  for (b).}
  \label{fig:sim1}
 \end{center}
\end{figure}

\vspace{12pt}
\begin{flushleft}
{\bf Zero temperature}
\end{flushleft}
We start with the Euclidean action:
\begin{equation}
 \frac{S_{\rm E}}{\hbar} = \frac{N}{2} \int_{-\infty}^{\infty} d\tau 
  \left( \frac{1}{2} \left(\frac{d\bvec{X}}{d\tau}\right) ^2
   + U (\bvec{X})\right), \label{action}
\end{equation}
where $U(\bvec{X})$ is the potential given by (\ref{ux}).
The bounce solution is the classical solution to the equations of motion
\begin{eqnarray}
 \frac{\mathrm{d} ^2}{\mathrm{d} \tau ^2} X - X + \frac{1}{X^3} 
  + P\frac{1}{X^2 YZ} &=& 0, \\
 \frac{\mathrm{d} ^2}{\mathrm{d} \tau ^2} Y - Y + \frac{1}{Y^3} 
  + P\frac{1}{XY^2 Z} &=& 0, \\
 \frac{\mathrm{d} ^2}{\mathrm{d} \tau ^2} Z - \lambda ^2Z + \frac{1}{Z^3} 
  + P\frac{1}{XYZ^2 } &=& 0,
\end{eqnarray}
subject to the boundary condition
\begin{equation}
 \lim_{\tau\to\pm\infty} \bvec{X} (\tau) = \bvec{X}_s \quad 
  (\mbox{stable critical point}).
\end{equation}
In Fig.\ref{fig:sim1}.
we show the behavior of bounce solutions obtained
using numerical simulations.
%
%

Let us investigate analytically 
the system (\ref{action}) by choosing a parameter $P$ near the 
critical value $P^*$. 
Then the bounce solution $\bvec{X}_b(\tau)$ is restricted 
in the neighborhood of the stable critical point $\bvec{X}_s$.
Indeed, the equation (\ref{crit}) gives the estimation,
$|\bvec{X}_b(\tau)-\bvec{X}_s| \sim |\bvec{X}_s-\bvec{X}_u| \sim 
{\cal O}((1-P/P^*)^{1/2})$.
In the following text we will assume 
\begin{eqnarray}
  \delta = 1- P/P^* \sim 10^{-3}, \quad |P| < |P^*|.
\end{eqnarray}
This parameter region is particularly interesting; 
as seen later on the value
$S_{\rm E}/\hbar$ is of the order one, in this region, though the
prefactor $N$ in the action is very large 
(the number of atoms
used in the experiment at Rice University is 
of order $10^3$~\cite{B-S-H, S-B-W-H}).
Thus we can expect to observe the macroscopic
quantum tunneling by experiments.

We now introduce a new coordinate $\bvec{x} = (x_T, x_N, x_B)$ 
around $\bvec{X}_s$
\begin{equation}
 \bvec{X} = \bvec{X}_s + \sum_{A=T,N,B} x_A \bvec{e}_A
\end{equation}
and expand the potential
\begin{equation}\label{potU}
 U(\bvec{X}) = U(\bvec{X}_s) + \frac{1}{2}\sum_{A=T,N,B} \epsilon_A^2 x_A^2 
+ \sum_{n+m+l = 3} c_{nml} x_T^n x_N^m x_B^l +\cdots .
\end{equation}
It should be noticed that the eigenvalue $\epsilon _T$ approaches to zero 
for $\delta \to 0$. Indeed, we can evaluate the behavior of
$\epsilon _T$ near $P^*$ using the exact formula (\ref{lambda-t}):
\begin{equation}\label{eigen}
 \epsilon _T = \alpha \delta^{1/4}+
  {\cal O}(\delta^{3/4}),
\end{equation}
where
\begin{equation}\label{alpha}
 \alpha^2 = \frac{4}{\lambda ^2 + 1 - P^*_{23}}
  \sqrt{(-6P^*_{23})(1-2P^*_{41})(2\lambda ^2 (1-P^*_{41}) -P^*_{23})}.
\end{equation}
On the other hand, eigenvalues $\epsilon _N$ and $\epsilon _B$ can be
approximated by (\ref{lambda-n}) and (\ref{lambda-b}) evaluated on
$P^*$, and these
values become extremely large compared with $\epsilon _T$ when the
parameter $\delta$ approaches to zero.
This means that the direction of the initial (infinitesimal) velocity of 
the bounce solution is given by the eigenfunction $\bvec {e} _T$.
Thus the trajectory of the bounce solution is mainly described by 
$x(\tau)= x_T(\tau)$, i.e. $T$-component of the coordinate
$\bvec{x}(\tau)$, 
and remaining components $x_N(\tau)$ and $x_B(\tau)$ give higher order
corrections. 
More precisely, using (\ref{potU}) and (\ref{eigen}), 
we can evaluate the bounce solution as 
$x_T(\tau) \sim {\cal O}(\delta^{1/2})$,
$x_N(\tau) \sim {\cal O}(\delta)$ and
$x_B(\tau) =0$ by the symmetry of the equations of motion 
(if we specialize to the spherically symmetric trapping potential,
the $N$-component $x_N(\tau)$ exactly vanishes).
We now approximate (\ref{action}) by one-dimensional 
quantum mechanical action:
\begin{equation}\label{Se}
 \frac{S_{\rm E}}{\hbar} \simeq \frac{N}{2} \int_{-\infty}^{\infty} d\tau 
  \left( \frac{1}{2} \left( \frac{dx}{dt}\right)^2 
+ \frac{1}{2} \epsilon_T ^2 x^2 + \frac{c}{3!} x^3 \right).
\end{equation}
From (\ref{eT}) and (\ref{T}), the coefficient $c(<0)$ is given by
\begin{equation}
 c =\left.
  \frac{1}{\triangle_T^3} 
  \left\{ 2T^3 \left( \frac{\partial ^3 U}{\partial X ^3} 
                           + 3 \frac{\partial ^3 U}{\partial X^2 \partial Y}\right)
   + P_{32}^3 \frac{\partial ^3 U}{\partial Z^3} 
   + 18 T^2 P_{32} \frac{\partial ^3 U}{\partial X \partial Y \partial Z}
   + 6 T P_{32}^2 \frac{\partial ^3 U}{\partial X  \partial Z^2}
                \right\} \right|_{\bvec{X} = \bvec{X}_s},
\end{equation}
which takes the following asymptotic form,
\begin{equation}\label{c}
 c = -12 (P_{24}^* + 4P_{41}^* P_{24}^* 
  - 2 \lambda^2 P_{42}^*)(1+2(P^*_{32})^2)^{-3/2} + 
{\cal O} ( \delta^{1/2}).
\end{equation}
It is convenient to introduce new scales characterizing the quantum
tunneling : according to Fig.3 we define
\begin{eqnarray}
&&
\mbox{(a) length scale} \quad R_0 = \frac{3a_0 \epsilon_T^2}{|c|} 
            =\frac{3a_0 \alpha^2}{|c|}\delta^{1/2}
            (1+{\cal O}(\delta^{1/2})),
            \label{R0} \\
&&
\mbox{(b) energy scale} \quad U_0 = \frac{N\hbar \nu \epsilon_T^6}{3c^2}
            =\frac{N\hbar \nu \alpha^6}{3c^2}\delta^{3/2}
            (1+{\cal O}(\delta^{1/2})). \label{U0}
\end{eqnarray}
Then we have a natural time scale
\begin{equation}\label{T0}
 T_0 = \frac{R_0}{\left( 2U_0/Nm \right)^{1/2}}
  = \frac{\omega _0}{\nu \alpha} 
  \delta^{-1/4}(1+{\cal O}(\delta^{1/2})), \quad 
  \omega _0 = \sqrt{\frac{27}{2}},
\end{equation}
representing the ``tunneling time''.
%
%
\begin{figure}[h]
\begin{center}
 \psfrag{Rm}[]{$R_m$}
 \psfrag{R0}[]{$R_0$}
 \psfrag{U0}[]{$U_0$}
 \psfrag{O}[]{$O$}
 \psfrag{R}[]{$R$}
 \psfrag{U}[]{$\hat{U}(R)$}
 \includegraphics[width=10cm,height=10cm,keepaspectratio,angle=-90]{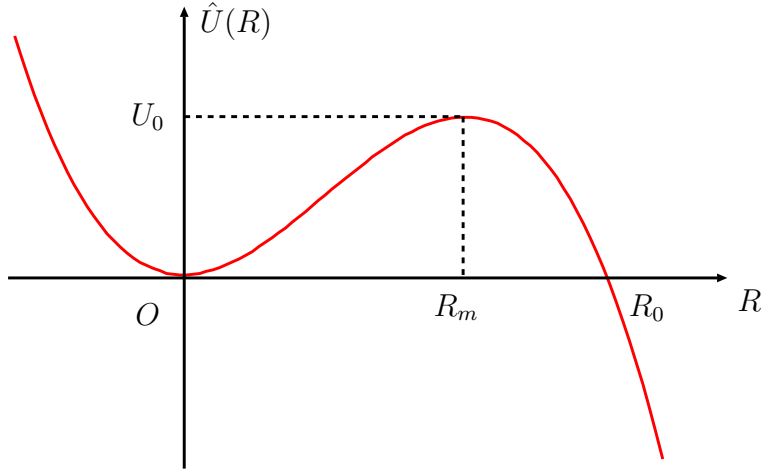}
 \caption{Potential profile for $\delta\rightarrow 0$. The potential 
 $\hat{U}(\bvec{W})$ given by (\ref{uhat}) is approximated by a
 one-dimensional potential $\hat{U}(R) = e_0 N (
 (\epsilon _T ^2/2) x^2 + (c/3!) x^3)$ with $R=a_0 x$. The potential 
$\hat{U}(R)$ has a metastable minimum at $R=0$ and a barrier of height 
$U_0 = \hat{U}(R_m),R_m=2 a_0 \epsilon _T ^2 / |c|$.}
 \label{fig:uw}
 \end{center}
\end{figure}

\begin{figure}[htbp]
 \begin{center}
  \subfigure[]{%
  \psfrag{0}[]{\scriptsize $0$}
  \psfrag{0.01}[]{\scriptsize $0.01$}
  \psfrag{0.1}[]{\scriptsize $0.1$}
  \psfrag{1}[]{\scriptsize $1$}
  \psfrag{10}[]{\scriptsize $10$}
  \psfrag{100}[]{\scriptsize $100$}
  \psfrag{20}[]{\scriptsize $20$}
  \psfrag{40}[]{\scriptsize $40$}
  \psfrag{60}[]{\scriptsize $60$}
  \psfrag{80}[]{\scriptsize $80$}
  \psfrag{100}[]{\scriptsize $100$}
  \psfrag{120}[]{\scriptsize $120$}
  \psfrag{140}[]{\scriptsize $140$}
  \psfrag{A}[]{\small $A(\lambda)$}
  \psfrag{B}[]{\small $B(\lambda)$}
  \psfrag{lambda}[]{\small $\lambda$}
  \includegraphics[width=8cm,height=8cm,keepaspectratio]{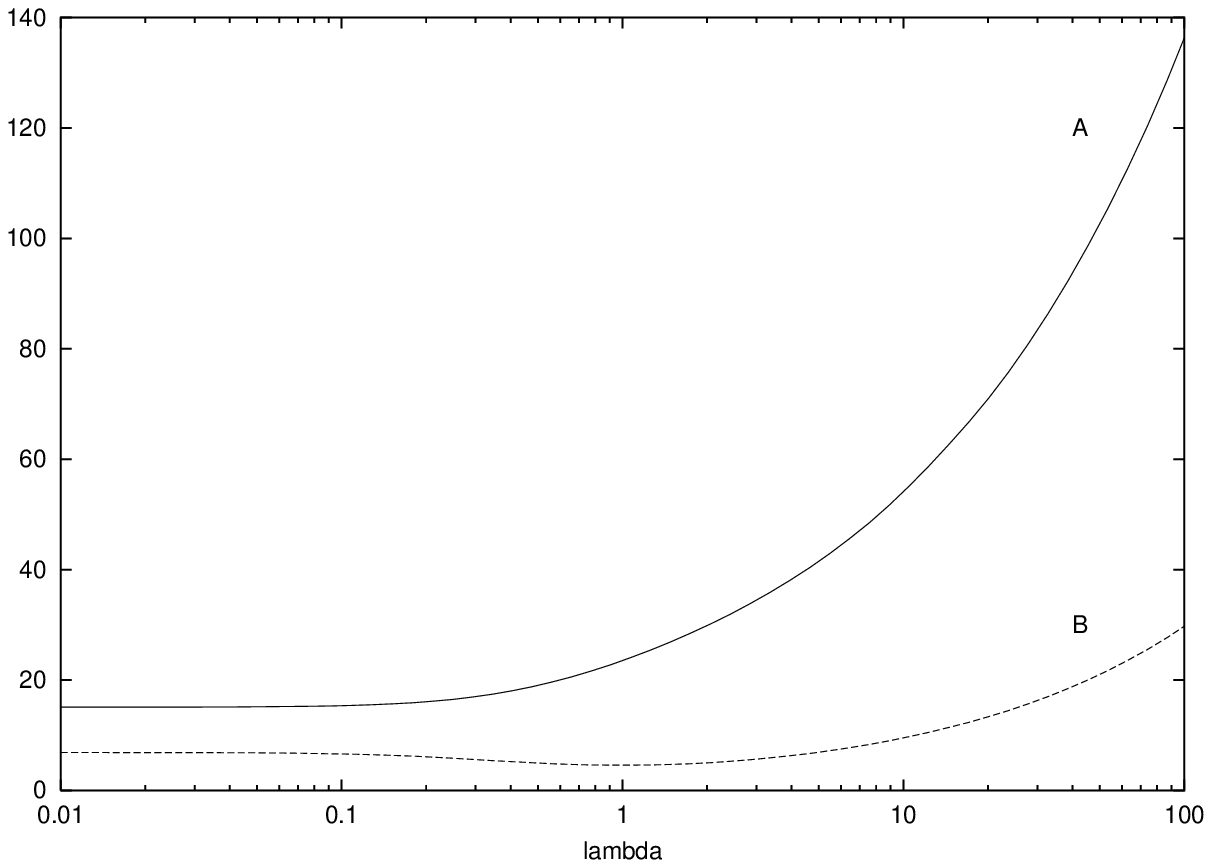}
  }
  \subfigure[]{
  \psfrag{0}[]{\scriptsize $0$}
  \psfrag{0.1}[]{\scriptsize $0.1$}
  \psfrag{1}[]{\scriptsize $1$}
  \psfrag{10}[]{\scriptsize $10$}
  \psfrag{2}[]{\scriptsize $2$}
  \psfrag{4}[]{\scriptsize $4$}
  \psfrag{6}[]{\scriptsize $6$}
  \psfrag{8}[]{\scriptsize $8$}
  \psfrag{10}[]{\scriptsize $10$}
  \psfrag{12}[]{\scriptsize $12$}
  \psfrag{14}[]{\scriptsize $14$}
  \psfrag{lambda}[]{\small $\lambda$}
  \psfrag{B}[]{\small $B(\lambda)$}
  \includegraphics[width=5cm,height=5cm,keepaspectratio]{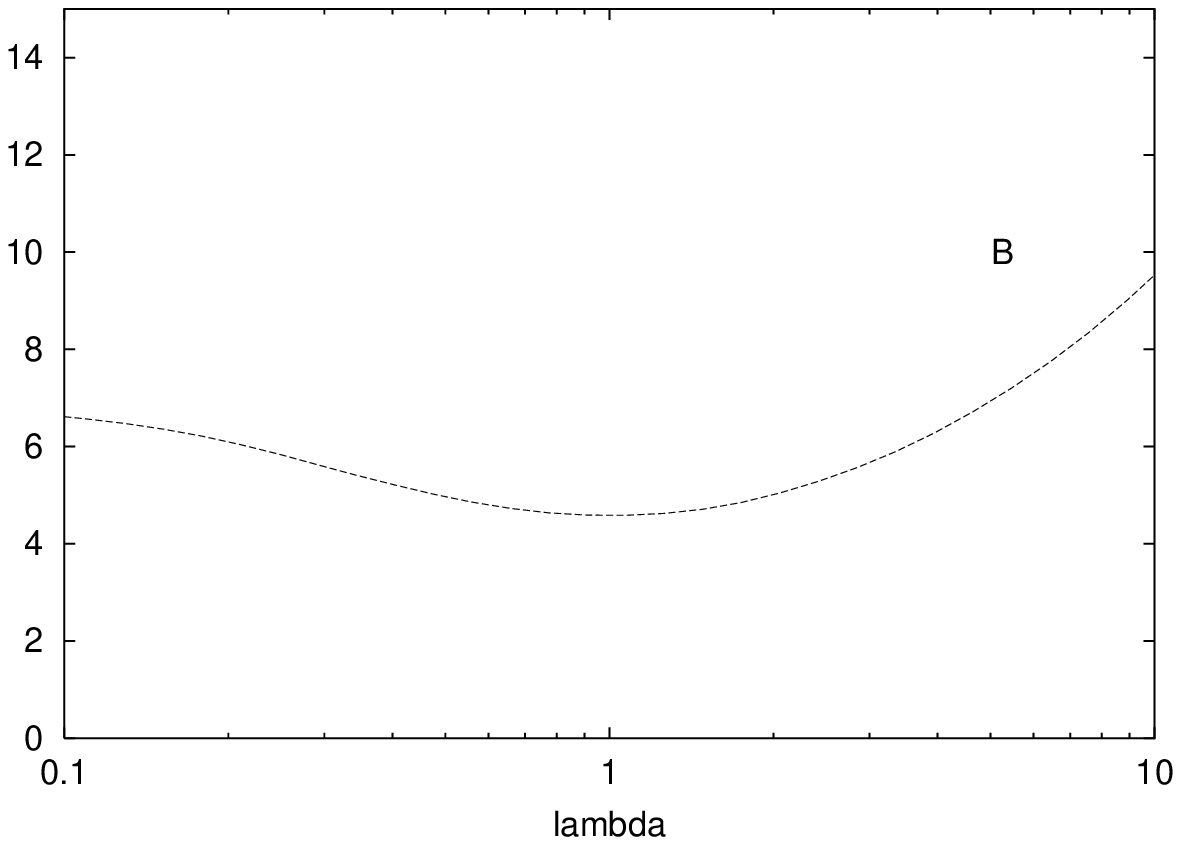}}
    \caption{$\lambda$-dependence of the functions $A$ and $B$. (b)
  shows the details of $B(\lambda)$ in the region of $\lambda  =1$.}
    \label{fig:ab}
 \end{center}
\end{figure}

Now the action (\ref{Se}) is of the form
\begin{equation}\label{faction}
 \frac{S_{\rm E}}{\hbar} = \frac{1}{h} \int_{-\infty}^{\infty} ds
  \left( \frac{1}{2} \left(\frac{dq}{ds}\right)^2+\tilde{U}(q) \right),
  \quad \tilde{U}(q)=\frac{1}{2} \omega _0 ^2 q^2 (1-q),
\end{equation}
where we have used rescaled quantities,
$q = (a_0/R_0) x$ and $s = (1/\nu T_0)\tau$.
The prefactor (effective Plank constant)
\begin{equation}\label{h}
 h = \frac{\hbar}{U_0 T_0} = \frac{2\omega _0 c^2}{9N\alpha^5}
 \delta^{-5/4}(1+{\cal O}(\delta^{1/2}))
\end{equation}
is a dimensionless parameter
controlling the validity of the WKB approximation .
The equations of motion can be easily integrated yielding the well known
bounce solution
\begin{equation}\label{sol}
q_b(s)={\rm sech}^2 \left(\frac{\omega_0 s}{2}\right).
\end{equation}
Using the WKB approximation~\cite{K},
we obtain the decay rate
\begin{equation}\label{Gamma0}
 \Gamma_0 = \left( 4\sqrt{\frac{\omega_0^3}{\pi h}}
                           \exp \left(-\frac{S_{\rm cl}}{h} \right)
			\right)   (1+{\cal O}(h)) 
		 {T_0}^{-1} 
\end{equation}
with the bounce action $S_{\rm cl} = \frac{8}{15} \omega _0$.
It follows from (\ref{T0}) and (\ref{h}) that for $\delta \to 0$
the leading contribution to $\Gamma_0$ is given by
\begin{eqnarray}
 \frac{\Gamma_0}{\nu} \simeq A \sqrt{N} \delta^{7/8}
 \exp \left(-BN \delta^{5/4}\right).
\end{eqnarray}
Here, the coefficients $A$ and $B$ are functions of the asymmetry
parameter $\lambda$:
\begin{eqnarray}
 A = 4 \sqrt{\frac{9}{2 \pi}} \frac{\alpha^{7/2}}{|c|}, \quad
 B = \frac{12 \alpha^5}{5c^2},
\end{eqnarray}
which can be calculated by (\ref{alpha}) and (\ref{c}).
Fig.\ref{fig:ab} shows the $\lambda$-dependence of these coefficients.
The spherically symmetric trapping potential $(\lambda =1)$
minimizes the function $B$ and its value is $4.58$ 
in excellent agreement with the result 
of~\cite{U-L}.
The functions $A,B$ remain relatively constant for $\lambda <1$ but they
grow for $\lambda >1$.
For $\delta \to 0$ the tunneling exponent and 
the prefactor vanish according to $\delta^{5/4}$ and $\delta^{7/8}$, 
respectively ~\cite{U-L, H-M-D-B-B}.
We find that this scaling law is universal,
independently of the shape of the harmonic trapping potential.

\vspace{12pt}
\begin{flushleft}
{\bf Finite temperature}
\end{flushleft}

\begin{figure}[h]
\begin{center}
 \psfrag{q0}[]{$q_0$}
 \psfrag{q1}[]{$q_1$}
 \psfrag{q2}[]{$q_2$}
 \psfrag{q}[]{$q$}
 \psfrag{-U}[]{$-\tilde{U}(q)$}
 \psfrag{O}[]{$O$}
 \psfrag{-1}[]{$-1$}
 \psfrag{-e}[]{$-E$}
 \includegraphics[width=10cm,height=10cm,keepaspectratio,angle=-90]{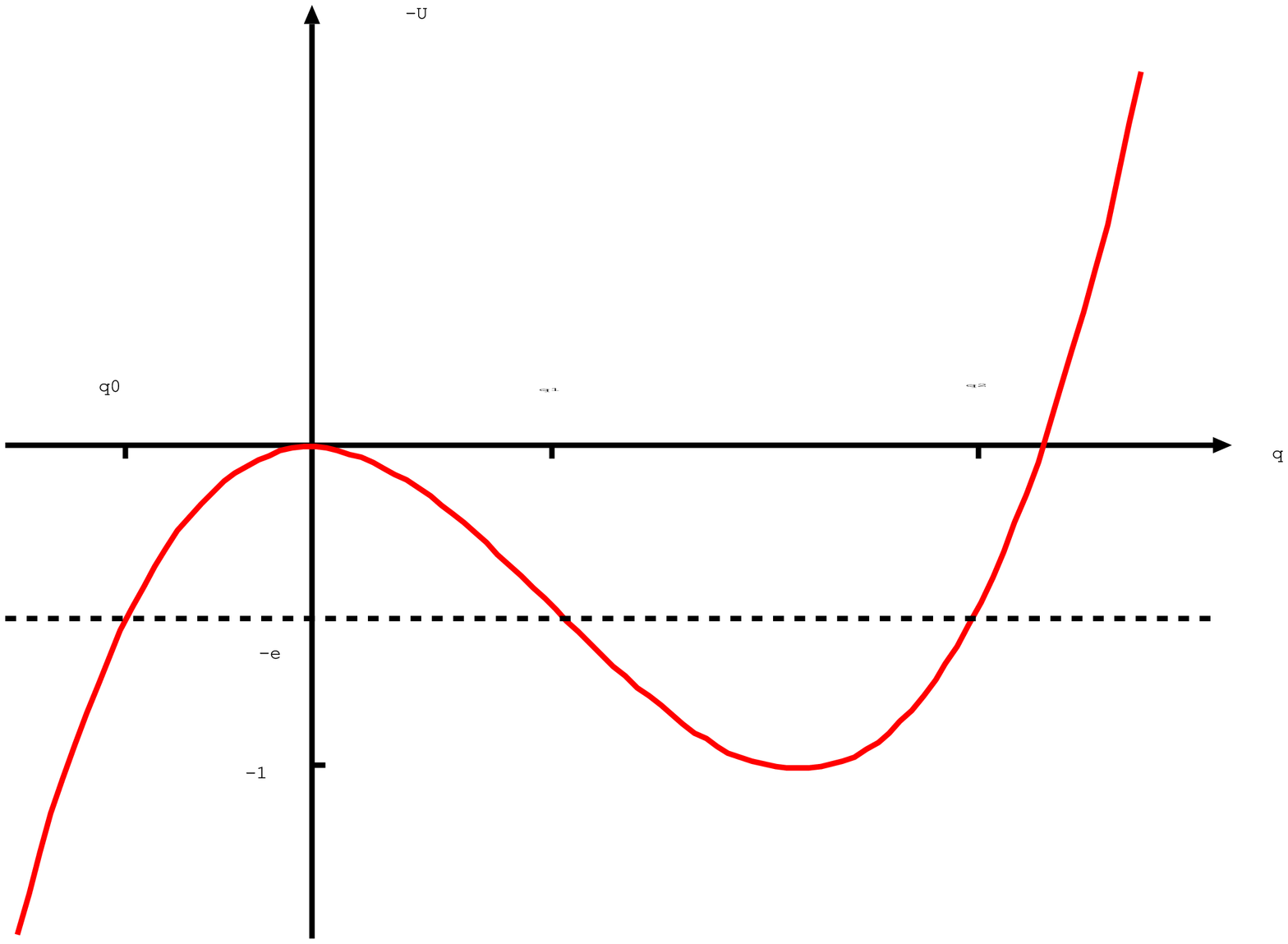}
 \caption{Turning points in the potential 
 $\tilde{U}(q) = \frac{1}{2} \omega _0 ^2 q^2 (1-q),\omega _0 = \sqrt{27/2}$. 
 The ``energy'' $E(0<E<1)$ is determined as a function of $\beta$ by
 requiring that the motion between the turning points $q_1$ and $q_2$ is
 periodic, with period $\beta h$.}
 \label{fig:uq}
 \end{center}
\end{figure}

In the case of finite temperature $\beta ^{-1}$, the bounce solution is
given by a periodic solution, i.e. the classical solution in the
potential $-\tilde{U}(q)$ with energy $-E \quad (0<E<1)$.
From Fig.\ref{fig:uq} the solution takes the form~\cite{Z},
\begin{equation}
 q_b(s) = q_2 - (q_2 - q_1) \mathrm{sn}  ^2 \left(\frac{\omega _0}{2}
          \sqrt{q_2 - q_0} s; m \right)
\end{equation}
with the elliptic modulus $m = \sqrt{\frac{q_2-q_1}{q_2-q_0}}$
and the period $h\beta$ is given
by the complete elliptic integral of the first kind :
\begin{equation}
 h\beta = \frac{4}{\omega _0 \sqrt{q_2 - q_0}} K(m), \quad
  K(m) = \int_0^1 \frac{dx}{\sqrt{(1-x^2)(1-mx^2)}}.
\end{equation}
%
%

This solution reduces, of course, to the previous solution
(\ref{sol}) for $E=0$.
The corresponding bounce action is evaluated as 
\begin{eqnarray}
 S_{\rm cl} &=& \int_0^{h\beta} ds 
  \left( \frac{1}{2} \left( \frac{dq_b}{ds}\right)^2 + \tilde{U}(q_b)\right)
  \nonumber \\
 &=& W + h\beta E, \label{W+E}
\end{eqnarray}
where
\begin{equation}
 W = \frac{4\omega_0}{15} \sqrt{q_2 - q_0}
  [2(q_0^2 + q_1^2 + q_2^2 - q_0q_1 - q_0q_2 - q_1q_2)E(m) + (q_1-q_0)(2q_0-q_1-q_2)K(m)]
\end{equation}
( $E(m)$ is the complete elliptic integral of the second kind ).
The fluctuation modes about the bounce solution include a zero mode 
$\phi_1(s) = \dot{q}_b(s)$. 
Then the determinant factor $A$ in (\ref{Gamma})
is calculated from the Gelfand-Yaglom formula~\cite{G-Y,K-C} :
\begin{equation}
 A(\beta) = \left. \frac{1}{\sqrt{\pi h}}
  \sqrt{\frac{\dot{\phi}_1(s)}{\dot{\phi}_2(s)}}
  \sinh{(\omega_0 s)}\right|_{s=\beta h /2},
 \quad 
 \phi_2(s)=\phi_1(s) \int^s \frac{ds'}{\phi_1(s')^2}.
\end{equation}
Thus we obtain the finite temperature decay rate due to quantum
tunneling : 
\begin{equation}\label{Gammabeta}
 \Gamma (\beta) = \left( A(\beta) \exp \left( \frac{-S_{\rm cl}}{h} \right) 
		   \right) (1+{\cal O}(h)) T_0^{-1},
\end{equation}
where
\begin{equation}\label{Abeta}
 A(\beta) = \sqrt{\frac{\omega_0^3}{2\pi h}} 
 \frac{(q_2-q_0)^{3/4} (q_2-q_1)(1-m^2)}{(a(m)E(m) + b(m)K(m))^{1/2}}
\sinh \left(\frac{\omega _0 \beta h}{2}\right) 
\end{equation}
with 
\begin{eqnarray}
 a(m) &=& 2(m^4 - m^2 + 1),\\
 b(m) &=& (1-m^2)(m^2 - 2).
\end{eqnarray}

For $E \to 0$, 
we have $(1-m^2)\sinh (\omega _0 \beta h/2) \to 8$,
$a(m)E(m)+b(m)K(m) \to 2$ and 
$q_0, q_1 \to 0, \, q_2 \to 1$, so that
$A(\beta) \to 4 \sqrt{\omega_0^3/\pi h}$,
which reproduces the zero-temperature decay rate $\Gamma_0$.
Let us turn now to the limit $E \to 1$,
where the period behaves as
\begin{equation}
 \beta h = \frac{2\pi}{\omega _0} \left(1 + \frac{5}{36} (1-E) + \cdots\right).
\end{equation}
The leading term gives a crossover temperature $\beta _c^{-1} =
h \omega _0/2\pi$~\cite{A}, i.e. for $\beta ^{-1} > \beta
_c^{-1}$ the decay rate is given by the familiar Arrhenuis-Kramers
formula~\cite{Kr}. 
On the
other hand, for $\beta ^{-1} < \beta _c^{-1}$ the macroscopic tunneling through the
barrier becomes more probable, and the decay rate is given by (\ref{Gammabeta}).
Recalling the energy unit $U_0$ defined by (\ref{U0}), and (\ref{h}) we find
\begin{equation}
 \beta _c^{-1} = \frac{h\omega _0}{2\pi} 
  \left(\frac{U_0}{k_B} \right)
  = \frac{\hbar\nu \alpha}{2\pi k_B} \delta^{1/4}(1+
  {\cal O}(\delta^{1/2})).
\end{equation}
For small $(\beta -\beta_c)/\beta_c > 0$, from (\ref{W+E}) and (\ref{Abeta}),
we obtain the
bounce action
\begin{eqnarray}\label{caction}
  \frac{S_{\rm cl}}{h} \simeq \beta - \frac{18}{5}\beta_c
  \left( \frac{\beta - \beta_c}{\beta_c}\right)^2,
\end{eqnarray}
and 
\begin{eqnarray}\label{aAbeta}
  A(\beta) \simeq \sqrt{\frac{8 \omega_0^3}{15h \pi^2}}
  \sinh{\left(\frac{\omega_0 \beta h}{2}\right)}
  \left(1-\frac{77}{20}\left( \frac{\beta - \beta_c}{\beta_c}\right)
  + \frac{20867}{2400} \left( \frac{\beta - \beta_c}{\beta_c}\right)^2
\right).
\end{eqnarray}


\begin{figure}[h]
\begin{center}
 \psfrag{lambda}[]{$\lambda$}
 \psfrag{gamma}[]{$\Gamma _0 \quad \mathrm{[sec^{-1}]}$}
 \includegraphics[width=10cm,height=10cm,keepaspectratio]{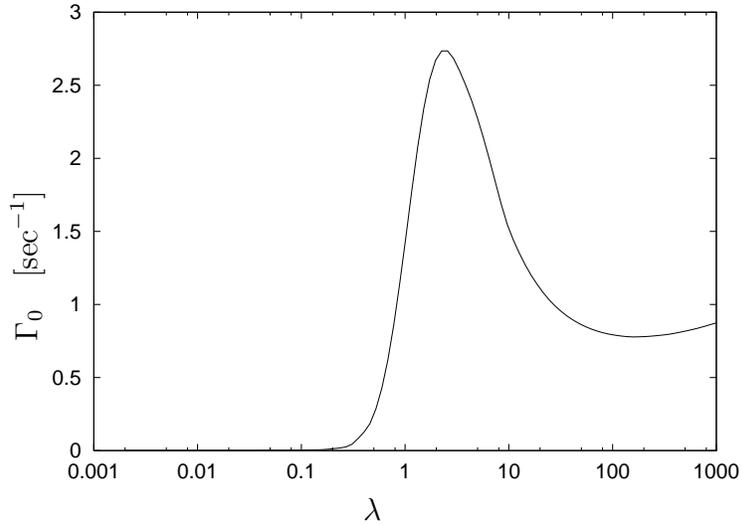}
 \caption{The decay rate $\Gamma _0$ as a function of the asymmetry
 parameter $\lambda$ for $\nu = 953 ~\mathrm{sec^{-1}}$, $a_0/a = -2.13 \times 10^3$
 and $\delta = 5.0 \times 10^{-3}$.}
 \label{fig:gamma005}
 \end{center}
\end{figure}


\section{Conclusion}
In this paper we have investigated the macroscopic tunneling of the
metastable condensate of $^7$Li.
When the number of particles in the condensate exceeds
a critical value $N^*=\sqrt{\pi/2}P^* a_0/a$,
the metastable condensate no longer exists and the equation 
giving critical points has no solutions.
In a region extremely close to $N^*$, i.e.
$\delta=1-P/P^* \ll 1$,
we have shown that the action takes a rather simple form (\ref{faction}),
and explicitly calculated the decay rate of the metastable condensate
using the WKB approximation.

Finally we make some remarks on our results.
In order to justify the WKB approximation, 
we should choose the effective Plank constant $h$ 
to satisfy the condition $h \ll 1$.
On the other hand, for very small $h$,
it is impossible to observe the macroscopic tunneling;
the formula (\ref{Gamma0}) provides an estimate of the tunneling decay rate,
$\Gamma_0 \sim {\cal O}({\rm e}^{-1/h})$.
This implies rather severe conditions on the parameter $\delta$
through the equation (\ref{h}).
If we use the experimental data at Rice University for the trapping
potential \cite{B-S-H,S-B-W-H};
$\lambda \simeq 0.867$, $a_0/a \simeq -2.13 \times 10^3$
and $\nu \simeq 953~{\rm sec}^{-1}$,
then the conditions are given by $h \simeq 2.92 \times 10^{-4}~ 
\delta^{-5/4} \ll 1$
and $\Gamma_0 \simeq 8.17 \times 10^5~\delta^{7/8} 
\exp{(- 6.72 \times 10^3~\delta^{5/4})} \sim 
{\cal O}(1)~{\rm sec}^{-1}$.
Consequently, we have a typical region $3.0 \times 10^{-3} < \delta <
7.0 \times 10^{-3}$.
Temperature effects on the tunneling decay rate are estimated
by using the equations (\ref{caction}) and (\ref{aAbeta}):
$\Gamma(\beta)$ is monotone decreasing for $\beta > \beta_c$
and hence $\Delta \Gamma =
\Gamma(\beta)-\Gamma_0 < \Gamma(\beta_c)-\Gamma_0$.
For instance, for $\delta = 5.0 \times 10^{-3}$ the decay rate at zero temperature
is $\Gamma_0 \simeq 1.03~{\rm sec}^{-1}$
and $\Delta \Gamma < 2.79~{\rm sec}^{-1}$.
The crossover temperature is then given by 
$\beta_c^{-1} \simeq 1.02~{\rm nK}$,
which may be a realizable temperature in the experiments.
The details of a crossover region have been discussed in~\cite{A};
there is a narrow crossover region of ${\cal O} (h^{3/2})$, where the
decay rate is given by
\begin{equation}
 \Gamma (\beta) T_0 \simeq \sqrt{\frac{8 \omega _0 ^3}{15 h \pi ^2}} 
  \sinh \left( \frac{\omega _0 \beta h}{2} \right)
		 \mathrm{erf} \left[ \sqrt{\frac{36}{5 \beta _c}} (\beta - \beta _c)
					 \right]
		 \exp \left[-\beta + \frac{18 \beta _c}{5} 
			   \left(\frac{\beta - \beta _c}{\beta _c}\right)^2
			 \right],
\end{equation}
with $\mathrm{erf}(x) = (2\pi) ^{-1/2} \int_{-\infty}^x dy \exp
(-y^2/2)$.
For very small $h\ll 10^{-2}$, this formula matches smoothly onto
(\ref{Gammabeta}) and $\Gamma (\beta) T_0 = (\omega _0/2\pi) [\sinh
(\omega _0 \beta h/2)/\sin (\omega _0\beta h/2)]\exp(-\beta)$ (Arrhenuis
- Kramers formula) near $\beta _c$.
However, we can not apply the formula to the macroscopic
tunneling since the value of $h$ in our situation is too large.
We leave the issue of crossover region for future research.
The shape of the trapping potential also has some effect on the 
behavior of the decay rate $\Gamma _0$:
as shown in Fig.\ref{fig:gamma005},
the effect is significant for the disk-shaped potential 
$(\lambda >1)$,
although it is rather small for the cigar-shaped potential
$(\lambda \ll 1)$ and $\Gamma_0$ is of order $10^{-3}~{\rm sec}^{-1}$,
independently of $\lambda$.



\begin{thebibliography}{99}

\bibitem{D-G-P-S}F. Dalfovo, S. Giorgini, L.P. Pitaevskii and
	S. Stringari
1999 {\it Rev. Mod. Phys.} {\bf 71} 463

\bibitem{G}E.P. Gross 1961 {\it Nuovo Cimento} {\bf 20} 454

\bibitem{P}L.P. Pitaevskii 1961 {\it Sov. Phys. JETP} {\bf 13} 451

\bibitem{M-S-H-V}A.J. Moerdijk, W.C. Stwalley, R.G. Hulet and
B.J. Verhaar 1994 {\it Phys. Rev. Lett.} {\bf 72} 40

\bibitem{A-Mc-S-H}E.R.I. Abraham, W.I. McAlexander, C.A. Sackett and
	R.G. Hulet
1995 {\it Phys. Rev. Lett.} {\bf 74} 1315

\bibitem{P-M-C-L-Z}V.M. P\'erez-Garc\'{\i}a, H. Michinel, J.I. Cirac,
M. Lewenstein and P. Zoller 1996 {\it Phys. Rev. Lett.} {\bf 77} 5320;
1997 {\it Phys. Rev.} A {\bf 56} 1424

\bibitem{S}H.T.C. Stoof 1997 {\it J. Stat. Phys.} {\bf 87} 1353

\bibitem{P-S-R}A. Parola, L. Salasnich and L. Reatto 1998 
{\it Phys. Rev.} A {\bf 57} R3180

\bibitem{U-L}M. Ueda and A.J. Leggett 1998 {\it Phys. Rev. Lett.} {\bf 80} 1576

\bibitem{H-M-D-B-B}C. Huepe, S. M\'etens, G. Dewel, P. Borckmans and
	M.E. Brachet 1999 {\it Phys. Rev. Lett.} {\bf 82} 1616

\bibitem{B-S-H}C.C. Bradley, C.A. Sackett and R.G. Hulet 1997
{\it Phys. Rev. Lett.} {\bf 78} 985

\bibitem{S-B-W-H}C.A. Sackett, C.C. Bradley, M. Welling and 
R.G. Hulet 1997 {\it Appl. Phys.} B {\bf 65} 433

\bibitem{K}H. Kleinert 1995 {Path Integrals in Quantum Mechanics
	Statistics and Polymer Physics} (World Scientific)

\bibitem{Z}W. Zweger 1983 {\it Z. Phys.} B {\bf 51} 301

\bibitem{G-Y}I.M. Gelfand and A.M. Yaglom 1960 {\it J. Math. Phys.}
{\bf 1} 48

\bibitem{K-C}H. Kleinert and A. Chervyakov 1998 {\it Phys. Lett.} A {\bf
	245} 345

\bibitem{A}I. Affleck 1980 {\it Phys. Rev. Lett.} {\bf 46} 388

\bibitem{Kr}H. Kramers 1990 {\it Physica} {\bf 7} 284
\end{thebibliography}
\end{document}